# Fuzzy-based Robust Precision Consensus Tracking for Uncertain Networked Systems with Cooperative-Antagonistic Interactions

*Amorey Lewis*

*Abstract*—In bipartite consensus tracking (BCT) tasks for nonlinear multiagent systems, stochastic disturbances and actuator faults are regarded as essential factors that hamper effective controller formulation and tracking precision improvement. To address these difficulties, we design an improved finite-time performance function (FTPF) for a fuzzy fault-tolerant distributed cooperative control scheme to achieve finite-time robust precision BCT tasks for nonlinear multiagent systems. The parameter selection range of the improved FTPF is relaxed, which renders systems to achieve better transient performance. Benefitting from stochastic Lyapunov stability theory, it is shown that all signals of systems are semi-global uniformly ultimately bounded in probability, and bipartite consensus errors can satisfy the arbitrary precision with probability in the predefined time. Finally, to verify its effectiveness, the proposed control scheme is applied to BCT tasks of a group of vehicles, which manifests anticipated control performance under various uncertainties.

## I. INTRODUCTION

IN the past few years, the research concerning cooperative control for multiagent systems (MASs) has attracted considerable attention due to its numerous potential applications in many fields. A s a p rimary r esearch t opic o f cooperative control, consensus requires some control variables (such as displacement and velocity) of a group of agents to reach an agreement. Following this requirement, the works [1]–[4] have proposed various control schemes to achieve the consensus of MASs. To name a few, Valcher *et al.* [1] presented a consensus control approach for homogeneous MASs with switching communication topology. According to the Internal Model Principle, a consensus control scheme for heterogeneous MASs suffering from input constraints was proposed in [2].

A common feature of the above consensus schemes is that they are designed through interactive cooperation among agents. In other words, in consensus results [1]–[4], topologies describing communications among agents exist only with non-negative edge weights. However, it is undeniable that antagonistic relations exist in some real-world scenarios. For instance, in a two-party political system, two parties may be antagonistic due to disagreements. Inspired by this fact, the bipartite consensus tracking (BCT) problem was investigated in works [5]–[9], which demands that all agents agree on some variables with the same modulus but different signs. For instance, Wen *et al.* [5] investigated the distributed BCT problem for linear MASs with a dynamic leader. In [6], the BCT problem for second-order MASs was studied, and the time of bipartite convergence can be predefined. Since stochastic disturbances are hard to avoid in the practical environment, numerous BCT results for stochastic MASs were presented in [10]–[12]. For example, Wu *et al.* [10] proposed a finite-time bipartite consensus protocol for stochastic MASs. The BCT problem for stochastic MASs with input saturation was investigated in work [11]. However, the above BCT results were obtained with actuators in normal operation. In fact, this situation may not be satisfied during systems operation. As an essential part of systems, actuators may undergo faults due to electromagnetic interference or air turbulence [13]. Actuator faults may have a detrimental effect on the tracking performance of MASs. It may even render desired control objectives challenging to achieve. In such a case, it may be hard to complete precise BCT tasks by exploiting existing control methods, especially for MASs subjected to stochastic disturbances.

As an optimization-based control approach, the prescribed performance control (PPC) in [14]–[20] has attracted considerable attention due to its ability to improve system performance. For instance, in [14], the tracking error of the considered system was constrained within specified boundaries by utilizing the PPC approach. Of note is that the above results [14]–[20] primarily focus on the steady-state performance of the system. Nevertheless, in some practical engineering [21]–[23], such as in the attitude tracking mission for the rigid spacecraft, the transient performance of attitude tracking is also critical, such that the rigid spacecraft can track a predefined attitude trajectory in a short time. In such a case, the finite-time performance function (FTPF) has been proposed in [24]–[28] to improve the convergence time of the conventional PPC method. To name a few, by designing an exponential FTPF,

Liu *et al.* [25] investigated the tracking control problem for non-strict feedback systems with the finite-time prescribed performance. In [26], a fault-tolerant tracking control strategy based on the FTPF was proposed for a class of strict feedback systems. However, it should be emphasized that the parameter selection range of the FTPF in [25]–[28] is tight, which hinders the application of FTPF, and it is even challenging to ensure that the system achieves the desired performance. Therefore, how to relax the parameter selection range and develop a fuzzy fault-tolerant distributed cooperative control scheme based on the FTPF to achieve finite-time robust precision BCT tasks for stochastic nonlinear MASs, which motivates the current research.

In this paper, the finite-time robust precision BCT problem is considered for stochastic nonlinear MASs subjected to actuator faults. The main contributions are listed as follows:

1) Different from existing finite-time BCT instances [8]–[10], the prescribed settling time for bipartite consensus errors is independent of the parameters of controllers and the initial states of MASs, which allows the settling time to be directly preset based on tasks requirements.

2) An improved FTPF is designed to program bipartite consensus errors of MASs. Compared to the existing results [25]–[28], the parameter selection range of the designed FTPF is relaxed, which enables systems to achieve better transient performance.

3) The proposed distributed fault-tolerant control scheme suppresses the influence of actuator faults and assorted uncertainties on systems performance while avoiding the over-parameterization problem in adaptive control methods, thereby improving the robustness of systems and reducing the redundancy of controller parameters.

The rest of this paper is organized as follows: Section II presents some preliminary knowledge, including the construction of the new FTPF. Section III shows the problem formulation for BCT tasks of stochastic MASs suffering from actuator faults. The designed procedure for the fault-tolerant controller is presented in Section IV. The stability analysis process is presented in Section V. Section VI provides simulation results of BCT tasks for second-order stochastic MASs and a group of vehicles. Conclusions of this paper can get in Section VII.

**Notations:** In this paper, $R^n$ and $R^{n \times r}$ represent the $n$-dimensional real space and the $n \times r$ real matrix space, respectively. $\mathbf{0}$ denotes a zero matrix with appropriate dimensions. $\varnothing$ is the empty set. For a real number $\psi$, $|\psi|$ denotes its absolute value. For a vector $\upsilon$, $\|\upsilon\|$ denotes its Euclidean norm. $e$ is the natural constant. $\mathrm{sgn}(\cdot)$ is a signum function. $\tanh(\cdot)$ denotes the hyperbolic tangent function. $\mathcal{C}^2$ denotes the set of all functions with continuous second partial derivatives. If there exists a continuous and strictly increasing function $\bar{\alpha}(x)$ that satisfies the conditions $\bar{\alpha}(0) = 0$ and $\lim_{x \to +\infty} \bar{\alpha}(x) = +\infty$, then the function $\bar{\alpha}(x)$ is called the $K_\infty$ function.

## II. PRELIMINARY KNOWLEDGE

### A. Graph Theory

A signed digraph $\mathscr{G} = \{\mathbb{V}, \mathbb{E}, A\}$ is used to describe communication relationships among agents, where $\mathbb{V} =$ $\{v_1, v_2, \ldots, v_N\}$ denotes the set of nodes, $\mathbb{E} \subseteq \mathbb{V} \times \mathbb{V} = \{(v_m, v_i) : v_m, v_i \in \mathbb{V}\}$ represents the set of edges, and $A = [a_{im}] \in R^{N \times N}$ is the weighted matrix with signed weights such that $a_{im} \neq 0 \Leftrightarrow (v_m, v_i) \in \mathbb{E}$ and $a_{im} = 0$, otherwise. Specifically, the edge $(v_m, v_i)$ means that the $i$-th agent can receive information from the $m$-th agent, and the $m$-th agent is said to be a neighbor of the $i$-th agent. If the weight $a_{im} > 0$, then there is a cooperative relationship between agents $i$ and $m$. Conversely, if the weight $a_{im} < 0$, this indicates an antagonistic relationship between agents $i$ and $m$. In this paper, suppose that the digraph $\mathscr{G}$ has no self-loops (i.e., $a_{ii} = 0$ or $(v_i, v_i) \notin \mathbb{E}, i = 1, \ldots, N$). $L_{\mathscr{G}} = D - A$ denotes the Laplacian matrix of the digraph $\mathscr{G}$, where $D = \mathrm{diag}\{\Sigma_{m=1}^N |a_{1m}|, \ldots, \Sigma_{m=1}^N |a_{Nm}|\}$ denotes the in-degree matrix. If a particular node $v_i$ can connect each node in the set of nodes $\mathbb{V}$ through directed paths, the digraph $\mathscr{G}$ contains a spanning tree. The particular node $v_i$ is said to be the root of the spanning tree.

In MASs, the leader unidirectionally transmits information to followers. In the communication topology of MASs, $b_i$ represents the weight between the $i$-th follower and the leader. Moreover, define the diagonal matrix $B_{\mathscr{G}} = \mathrm{diag}\{|b_1|, \ldots, |b_N|\}$, where $b_i \neq 0$ if the $i$-th follower can communicate directly with the leader. Otherwise, $b_i = 0$. In this paper, the leader signal and its first derivative are bounded and they can be received by followers.

*Definition 1:* [5] The signed digraph $\mathscr{G}$ is called structurally balanced if the set of nodes $\mathbb{V}$ can provide a partition $\{\mathbb{V}_1, \mathbb{V}_2\}$ satisfying conditions $\mathbb{V}_1 \cup \mathbb{V}_2 = \mathbb{V}$ and $\mathbb{V}_1 \cap \mathbb{V}_2 = \varnothing$. Moreover, if nodes $v_i$ and $v_m$ exist in the same subset $\mathbb{V}_1$ or $\mathbb{V}_2$, then $a_{im} > 0$. Otherwise, $a_{im} < 0$.

*Assumption 1:* The considered communication topology of MASs contains a spanning tree, and the leader is the root of the spanning tree. Meanwhile, the communication topology is structurally balanced.

*Lemma 1:* [11] Under Assumption 1, the matrix $(L_{\mathscr{G}} + B_{\mathscr{G}})$ is nonsingular.

*Lemma 2:* [5] Define a set of diagonal matrices $\check{\mathcal{S}} = \{S = \mathrm{diag}\{\Im_1, \Im_2, \ldots, \Im_N\}, \Im_i \in \{\pm 1\}\}$. If Assumption 1 is valid, then there exists a matrix $S \in \check{\mathcal{S}}$ such that the matrix $SAS$ has all nonnegative entries. Moreover, the matrix $S$ gives a partition, i.e. $\mathbb{V}_1 = \{v_i | \Im_i > 0\}$ and $\mathbb{V}_2 = \{v_i | \Im_i < 0\}$.

### B. Stochastic Stability Theorem

The considered stochastic system is modeled as the following Itô-type stochastic differential equation:

$$dx = f(x)dt + g(x)d\varpi \tag{1}$$

where the vector $x \in R^n$ represents the system state, and $\varpi \in R^r$ denotes an independent standard Brownian motion. The functions $f(\cdot) : R^n \to R^n$ and $g(\cdot) : R^n \to R^{n \times r}$ satisfy the local Lipschitz condition with $f(\mathbf{0}) = \mathbf{0}$ and $g(\mathbf{0}) = \mathbf{0}$.

*Definition 2:* [29] For the system (1), if there exists a Lyapunov function $V(x) \in \mathcal{C}^2$, then the infinitesimal generator $\mathcal{L}$ of the function $V(x)$ is defined as

$$\mathcal{L}V(x) = \frac{\partial V(x)}{\partial x} f(x) + \frac{1}{2} Tr\{G$$

where $G = g^T(x)\frac{\partial^2 V(x)}{\partial x^2}g(x)$, and $Tr\{G\}$ represents the trace of the matrix $G$.

*Lemma 3:* [11] It is assumed that there exists a Lyapunov function $V(x) \in \mathcal{C}^2$ satisfying the following relationships

$$\mathcal{H}_1(\|x\|) \le V(x) \le \mathcal{H}_2(\|x\|)$$
$$\mathcal{L}V(x) \le -\varkappa_1 V(x) + \varkappa_2$$

where $\mathcal{H}_1(\|x\|)$ and $\mathcal{H}_2(\|x\|)$ belong to the $K_\infty$ function. $\varkappa_1$ and $\varkappa_2$ are positive constants. Then the stochastic system (1) is said to be semi-global uniformly ultimately bounded (SGUUB) in probability. Moreover, the solution of Eq. (1) satisfies the following condition

$$E[V(x)] \le V(x(0))e^{-\varkappa_1 t} + \frac{\varkappa_2}{\varkappa_1}, \quad \forall t > 0$$

where $x(0)$ is the initial state, and $E(\cdot)$ represents the mathematical expectation.

### C. Finite-Time Performance Function

*Definition 3:* [30] A function $\sigma(t)$ is said to be an FTPF if the function $\sigma(t)$ possesses the following properties

1. $\sigma(t)$ is a smooth function.
2. For $\forall t > 0$, $\sigma(t)$ satisfies $\sigma(t) > 0$ and $\dot{\sigma}(t) \le 0$.
3. $\lim_{t \to T_s} \sigma(t) = \sigma_\infty > 0$, where $\sigma_\infty$ is a constant.
4. For $\forall t \ge T_s$, $\sigma(t) = \sigma_\infty$, where $T_s$ is the settling time.

*Lemma 4:* For the given constants $\varsigma \ge 1$ and $\sigma_0 > \sigma_\infty > 0$, the following segmentation function is an FTPF.

$$\sigma(t) = \begin{cases} (\sigma_0 - \sigma_\infty)e^{\varsigma(1 - \frac{T_s}{T_s - t})} + \sigma_\infty, & 0 \le t < T_s \\ \sigma_\infty, & t \ge T_s \end{cases}$$

where $T_s$ represents the settling time, and it depicts the convergence rate of the FTPF. Moreover, $\sigma_\infty$ can be arbitrarily small, and it denotes the ultimate boundary of the FTPF.

*Proof:* Firstly, we give the proof of the first property.

Case 1: If $t \ge T_s$, then $\frac{d^\tau \sigma(t)}{dt^\tau} = 0$ with $\tau \ge 1$. Therefore, it is easily determined that $\frac{d^\tau \sigma(t)}{dt^\tau}$ are continuous and $\lim_{t \to T_s^+} \frac{d^\tau \sigma(t)}{dt^\tau} = 0$.

Case 2: If $0 \le t < T_s$, then $\sigma(t) = (\sigma_0 - \sigma_\infty)e^{\alpha(t)} + \sigma_\infty$, where $\alpha(t) = \varsigma(1 - \frac{T_s}{T_s - t}) = \frac{\varsigma t}{t - T_s}$.

Taking the first-order derivative of $\sigma(t)$ with respect to time $t$, we can obtain

$$\frac{d\sigma(t)}{dt} = (\sigma_0 - \sigma_\infty)\frac{d\alpha(t)}{dt}e^{\alpha(t)} \qquad (2)$$

where $\frac{d\alpha(t)}{dt} = \frac{-\varsigma T_s}{(t - T_s)^2}$. With the help of the L'Hospital's rule, it is not difficult to obtain

$$\lim_{t \to T_s^-} (\sigma_0 - \sigma_\infty)\frac{d\alpha(t)}{dt}e^{\alpha(t)}$$
$$= \lim_{t \to T_s^-} (\sigma_0 - \sigma_\infty)\left[\frac{\frac{-2}{\varsigma T_s}}{e^{-\frac{\varsigma T_s}{t - T_s}}}\right] = 0. \qquad (3)$$

Based on Case 1, Eqs. (2) and (3), it is derived that

$$\lim_{t \to T_s^+} \frac{d\sigma(t)}{dt} = \lim_{t \to T_s^-} \frac{d\sigma(t)}{dt} = 0$$

which indicates that the function $\frac{d\sigma(t)}{dt}$ is continuous, and $\sigma(t)$ is differentiable.

Based on Eq. (2), the second-order derivative of $\sigma(t)$ with respect to time $t$ can be further obtained by

$$\frac{d^2\sigma(t)}{dt^2} = (\sigma_0 - \sigma_\infty)\left[\frac{d^2\alpha(t)}{dt^2}e^{\alpha(t)} + \left(\frac{d\alpha(t)}{dt}\right)^2 e^{\alpha(t)}\right]. \qquad (4)$$

By applying the L'Hospital's rule, it yields

$$\lim_{t \to T_s^-} (\sigma_0 - \sigma_\infty)\frac{d^2\alpha(t)}{dt^2}e^{\alpha(t)}$$
$$= \lim_{t \to T_s^-} (\sigma_0 - \sigma_\infty)\left[\frac{\frac{-6}{(t - T_s)^2}}{e^{-\frac{\varsigma T_s}{t - T_s}}}\right] = 0. \qquad (5)$$

In addition, it can also derived that

$$\lim_{t \to T_s^-} (\sigma_0 - \sigma_\infty)\left(\frac{d\alpha(t)}{dt}\right)^2 e^{\alpha(t)}$$
$$= \lim_{t \to T_s^-} (\sigma_0 - \sigma_\infty)\left[\frac{\frac{-4\varsigma T_s}{(t - T_s)^3}}{e^{-\frac{\varsigma T_s}{t - T_s}}}\right] = 0. \qquad (6)$$

Based on Case 1, and Eqs. (4)-(6), one gets

$$\lim_{t \to T_s^+} \frac{d^2\sigma(t)}{dt^2} = \lim_{t \to T_s^-} \frac{d^2\sigma(t)}{dt^2} = 0$$

which indicates that $\frac{d^2\sigma(t)}{dt^2}$ is continuous, and $\sigma(t)$ being the second-order differentiable.

Based on the above analysis, it can deduce that $\frac{d^\tau\alpha(t)}{dt^\tau} = \frac{\hat{\gamma}_\tau}{(t - T_s)^{\tau+1}}$ and $\left(\frac{d\alpha(t)}{dt}\right)^\tau = \frac{\tilde{\gamma}_\tau}{(t - T_s)^{2\tau}}$, where $\hat{\gamma}_\tau$ and $\tilde{\gamma}_\tau$ are constants, and $3 \le \tau \le n - 1$. Therefore, $\frac{d^\tau\sigma(t)}{dt^\tau}$ can be denoted as a polynomial of $\frac{e^{\alpha(t)}}{(t - T_s)^m}$ with $m \ge 0$.

Through the above analysis, it can be obtained

$$\lim_{t \to T_s^-} \left[\frac{e^{\alpha(t)}}{(t - T_s)^m}\right] = -\frac{m}{\varsigma T_s}\lim_{t \to T_s^-}\left[\frac{\frac{1}{(t - T_s)^{m-1}}}{e^{-\frac{\varsigma T_s}{t - T_s}}}\right]$$
$$\cdots$$
$$= (-1)^{(m+1)}\frac{\prod_{j=0}^{(m-1)}(m-j)}{(\varsigma T_s)^{(m-1)}}\lim_{t \to T_s^-}\left[\frac{\frac{1}{t - T_s}}{e^{-\frac{\varsigma T_s}{t - T_s}}}\right] = 0. \qquad (7)$$

According to Case 1 and (7), it follows that

$$\lim_{t \to T_s^-}\left[\frac{d^\tau\sigma(t)}{dt^\tau}\right] = \lim_{t \to T_s^+}\left[\frac{d^\tau\sigma(t)}{dt^\tau}\right] = 0$$

which means that $\frac{d^\tau\sigma(t)}{dt^\tau}$ is continuous, and $\sigma(t)$ is $\tau$-times differentiable.

At last, by taking $\tau = n$, it yields

$$\lim_{t \to T_s^-}\left[\frac{d^n\sigma(t)}{dt^n}\right] = 0.$$

It should notice that

$$\lim_{t \to T_s^-}\left[\frac{d^n\sigma(t)}{dt^n}\right] = \lim_{t \to T_s^+}\left[\frac{d^n\sigma(t)}{dt^n}\right] = 0$$

which indicates that $\frac{d^n\sigma(t)}{dt^n}$ is continuous, and $\sigma(t)$ is $n$-times differentiable.

The above analysis can conclude that the function $\sigma(t)$ satisfies the first property. Furthermore, it is not difficult to verify that $\sigma(t)$ satisfies other properties. According to Definition 3, $\sigma(t)$ is an FTPF.

This completes the proof. ∎

*Remark 1:* In [25]–[28], the parameter selection range of the FTPF is tight, which hampers the application of FTPF, and it is even hard to guarantee that MASs achieve the desired performance. Compared to the existing results [25]–[28], the parameter selection range of the improved FTPF is relaxed, which renders systems to achieve better transient performance.

### D. Fuzzy Logic System

The fuzzy logic system (FLS) has excellent approximation ability to the unknown nonlinear function. The conventional fuzzy rules are as follows:

$R^\iota$: If $x_1$ is $F_1^\iota$, ..., and $x_n$ is $F_n^\iota$, then $y_F$ is $W^\iota$.

Specifically, $x = [x_1, \ldots, x_n]^T$ and $y_F$ are input and output of the FLS, respectively. $F_j^\iota$ and $W^\iota$ are fuzzy sets with membership functions $\phi_{F_j^\iota}(x_j)$ and $\phi_{W^\iota}(x_j)$, where $j = 1, 2, \ldots, n$ and $\iota = 1, 2, \ldots, P$. Based on the results in [31]–[39], the output $y_F$ of the FLS can be written as

$$y_F(x) = \frac{\sum_{\iota=1}^{P} \bar{y}_F^\iota \left[ \prod_{j=1}^{n} \phi_{F_j^\iota}(x_j) \right]}{\sum_{\iota=1}^{P} \left[ \prod_{j=1}^{n} \phi_{F_j^\iota}(x_j) \right]} \quad (8)$$

where $\bar{y}_F^\iota = \max_{y_F \in R}\{\phi_{W^\iota}(y_F)\}$. The fuzzy basis function can be represented by

$$\Phi_\iota(x) = \frac{\prod_{j=1}^{n} \phi_{F_j^\iota}(x_j)}{\sum_{\iota=1}^{P} \left[ \prod_{j=1}^{n} \phi_{F_j^\iota}(x_j) \right]}.$$

Denote $\theta = [\bar{y}_F^1, \bar{y}_F^2, \ldots, \bar{y}_F^P]^T = [\theta_1, \theta_2, \ldots, \theta_P]^T$ and $\Phi(x) = [\Phi_1(x), \Phi_2(x), \ldots, \Phi_P(x)]^T$. Eq. (8) can be further written as

$$y_F(x) = \theta^T \Phi(x).$$

*Lemma 5:* [40] If $r(x)$ is a continuous function defined on a compact set $\Omega_r$, then there exists an FLS $\theta^T \Phi(x)$ satisfying the following relationship

$$\sup_{x \in \Omega_r} |r(x) - \theta^T \Phi(x)| \leq \epsilon_r$$

where $\epsilon_r$ is any positive constant.

## III. PROBLEM FORMULATION

In this paper, the $i$-th follower is modeled as

$$\begin{cases} dx_{ij} = \left( x_{i(j \oplus 1)} + f_{ij}(\bar{x}_{ij}) \right) dt + g_{ij}(\bar{x}_{ij}) d\varpi \\ dx_{in_i} = \left( \sum_{h=1}^{M} l_{ih}\omega_{ih} + f_{in_i}(\bar{x}_{in_i}) \right) dt + g_{in_i}(\bar{x}_{in_i}) d\varpi \\ y_i = x_{i1}, \quad 1 \leq i \leq N, \quad 1 \leq j \leq n_i - 1 \end{cases} \quad (9)$$

where $\bar{x}_{i\ell} = [x_{i1}, x_{i2}, \ldots, x_{i\ell}]^T (\ell = 1, 2, \ldots, n_i)$ and $y_i \in R$ represent the state variables and output of the $i$-th follower, respectively. $f_{i\ell}(\cdot) : R^\ell \to R$ and $g_{i\ell}(\cdot) : R^\ell \to R^{1 \times r}$ denote unknown nonlinear functions satisfying the local Lipschitz condition. $f_{i\ell}(\mathbf{0}) = 0$ and $g_{i\ell}(\mathbf{0}) = \mathbf{0}$. $l_{ih} \in R$ is an unknown coefficient. $\omega_{ih}$ denotes the output of the $h$-th actuator of the $i$-th follower. $M$ represents the number of the actuator. $\varpi$ is an independent $r$-dimensional standard Brownian motion, which is defined on a complete probability space $(\bar{\Omega}, \mathscr{F}, \{\mathscr{F}_t\}_{t \geq 0}, \bar{P})$. $\bar{\Omega}$ stands for a sample space. $\mathscr{F}$ denotes a $\bar{\sigma}$-filed. $\{\mathscr{F}_t\}_{t \geq 0}$ represents a filtration, and $\bar{P}$ is a probability measure. In the subsequent derivation, for simplicity, functions $f_{i\ell}(\bar{x}_{i\ell})$ and $g_{i\ell}(\bar{x}_{i\ell})$ concerning states variables $\bar{x}_{i\ell}$ will be denoted by $f_{i\ell}$ and $g_{i\ell}$, respectively.

The actuator fault model is described by

$$\begin{cases} \omega_{ih}(t) = \rho_{ih}u_{ih}(t) + \nu_{ih}, \quad t \in [t_{ih\iota}^s, t_{ih\iota}^e) \\ \rho_{ih}\nu_{ih} = 0, \quad \iota = 1, 2, 3, \ldots \end{cases} \quad (10)$$

where $\rho_{ih} \in [0, 1]$ and $\nu_{ih}$ are unknown constants. $u_{ih}(t)$ is the input of the $h$-th actuator of the $i$-th follower. $\iota$ indicates the $\iota$-th actuator fault. $t_{ih\iota}^s$ and $t_{ih\iota}^e$ represent the moments when the fault occurs and ends, respectively. Eq. (10) involves the following three situations:

- If $\rho_{ih} = 1$ and $\nu_{ih} = 0$, then actuators work normally.
- For unknown constants $\underline{\rho}_{ih}$ and $\bar{\rho}_{ih}$, if $0 < \underline{\rho}_{ih} \leq \rho_{ih} \leq \bar{\rho}_{ih} < 1$ and $\nu_{ih} = 0$, then actuators undergo the partial loss of control effectiveness (PLOE).
- If $\rho_{ih} = 0$ and $\nu_{ih} \neq 0$, then actuators occur the total loss of effectiveness (TLOE).

*Control Objectives:* This paper aims at developing an FTPF-based control scheme for stochastic MASs to complete the following control objectives:

(a) All signals of MASs are SGUUB in probability.

(b) Bipartite consensus errors satisfy the arbitrary precision with probability in the predefined settling time.

Other assumptions and lemmas need to be provided to guarantee that control objectives are achieved.

*Assumption 2:* Only up to $M - 1$ actuators are allowed to work in the TLOE mode.

*Assumption 3:* For the dynamics model of the $i$-th follower, the signs of $l_{ih}$ $(h = 1, 2, \ldots, M)$ are known.

*Lemma 6:* [11] (Young's inequality): For $\forall \mathcal{P} \in R^n$, $\forall \mathcal{Q} \in R^n$, the following relationship holds

$$\mathcal{P}\mathcal{Q} \leq \frac{\varepsilon^{c_1}}{c_1}|\mathcal{P}|^{c_1} + \frac{1}{c_2\varepsilon^{c_2}}|\mathcal{Q}|^{c_2}$$

where $c_1 > 1$, $c_2 > 1$, $\varepsilon > 0$, and $\frac{1}{c_1} + \frac{1}{c_2} = 1$.

*Lemma 7:* [41] For any given variable $\Xi$, one has

$$0 \leq |\Xi| - \frac{\Xi^2}{\sqrt{\Xi^2 + \psi^2}} \leq \psi$$

where $\psi$ is a positive constant.

## IV. FAULT-TOLERANT CONTROLLER DESIGN

### A. Error Transformation

To achieve the control objective (b), the error transformation mechanism in [42] is adopted. Before giving the error transformation mechanism, the bipartite consensus error $z_{i1}$ is defined as

$$z_{i1} = \sum_{m=1}^{N} |a_{im}|(y_i - \text{sgn}(a_{im})y_m) + |b_i|(y_i - \text{sgn}(b_i)y_r) \quad (11)$$

where $y_r$ denotes the leader signal.

The error transformation mechanism is given as

$$z_{i1} = \sigma(t)\mu(e_{i1}^*), \quad |z_{i1}(0)| \in (0, \sigma(0)) \tag{12}$$

where $\mu(e_{i1}^*) = \frac{2}{\pi}\arctan(e_{i1}^*)$, and $e_{i1}^*$ is the transformed error. Based on Eq. (12) and the fourth property of the FTPF, for $\forall t \geq T_s$, one gets

$$-\sigma_\infty < z_{i1} < \sigma_\infty \tag{13}$$

which indicates that the bipartite consensus error $z_{i1}$ can converge to a predefined region in a finite time $T_s$. In the subsequent coordinate transformations, $e_{i1}^*$ will replace $z_{i1}$ to participate in the construction of the controller.

By taking the differential of Eqs. (11) and (12), one obtains

$$
\begin{aligned}
dz_{i1} &= \mu(e_{i1}^*)d\sigma(t) + \sigma(t)\frac{\partial\mu(e_{i1}^*)}{\partial e_{i1}^*}de_{i1}^* \\
&= \Big(q_i(x_{i2} + f_{i1}) - \sum_{m=1}^{N} a_{im}(x_{m2} + f_{m1}) - b_i\dot{y}_r\Big)dt \\
&\quad + \Big(q_i g_{i1} - \sum_{m=1}^{N} a_{im}g_{m1}\Big)d\varpi
\end{aligned} \tag{14}
$$

with $q_i = \sum_{m=1}^{N}|a_{im}| + |b_i|$.

Furthermore, it follows that

$$
\begin{aligned}
de_{i1}^* &= \xi_i\Big(q_i(x_{i2} + f_{i1}) - \sum_{m=1}^{N} a_{im}(x_{m2} + f_{m1}) - b_i\dot{y}_r - \beta_i\Big)dt \\
&\quad + \xi_i\Big(q_i g_{i1} - \sum_{m=1}^{N} a_{im}g_{m1}\Big)d\varpi
\end{aligned} \tag{15}
$$

where $\beta_i = \frac{2}{\pi}\arctan(e_{i1}^*)\dot{\sigma}(t)$, and $\xi_i = \frac{\pi(1+e_{i1}^{*2})}{2\sigma(t)}$.

*Remark 2:* It is necessary to ensure that bipartite consensus error $z_{i1}$ satisfies inequality (13) to achieve control objective (b). Therefore, the error transformation mechanism (12) is introduced to guarantee that inequality (13) holds. The mechanism (12) means that $z_{i1} \to \pm\sigma_\infty$ if and only if $e_{i1}^* \to \pm\infty$, for $\forall t \geq T_s$. Consequently, as long as $e_{i1}^*$ is bounded, inequality (13) can be guaranteed to be true, thus ensuring that control objective (b) can be achieved.

### B. Fault-Tolerant Controller Design

Before the design of the fault-tolerant controller, the coordinate transformations are provided as

$$
\begin{cases}
\zeta_{i1} = e_{i1}^* \\
\zeta_{i\hbar} = x_{i\hbar} - \alpha_{i(\hbar-1)}^*, \quad \hbar = 2, 3, \ldots n_i
\end{cases} \tag{16}
$$

where $\alpha_{i(\hbar-1)}^*$ is the output of the first-order command filter with respect to the virtual controller $\alpha_{i(\hbar-1)}$. $\alpha_{i(\hbar-1)}^*$ and $\alpha_{i(\hbar-1)}$ satisfy the following relationship

$$\tau_{i(\hbar-1)}\dot{\alpha}_{i(\hbar-1)}^* + \alpha_{i(\hbar-1)}^* = \alpha_{i(\hbar-1)}$$

where $\tau_{i(\hbar-1)}$ is a positive designed parameter, and initial values of $\alpha_{i(\hbar-1)}^*$ and $\alpha_{i(\hbar-1)}$ satisfy $\alpha_{i(\hbar-1)}^*(0) = \alpha_{i(\hbar-1)}(0)$.

**Step 1:** Choose the following Lyapunov function

$$V_{i1} = \frac{1}{4}\bar{\zeta}_{i1}^4 + \frac{1}{2\delta_i}\tilde{\Theta}_i^2$$

where $\bar{\zeta}_{i1} = \zeta_{i1} - \eta_{i1}$ is a compensated error, and $\tilde{\Theta}_i = \Theta_i - \hat{\Theta}_i$. $\Theta_i \triangleq \max_{1 \leq i \leq n_i}\{\|\theta_{i1}\|^{\frac{4}{3}}, \|\theta_{i2}\|^2\}$ is an unknown constant, and $\hat{\Theta}_i$ is the estimated value of $\Theta_i$. $\delta_i$ is a positive designed parameter. To compensate the filtering error $\alpha_{i1}^* - \alpha_{i1}$, the compensation signal $\eta_{i1}$ is designed as

$$
\begin{aligned}
\dot{\eta}_{i1} &= -(k_{i1}+1)\eta_{i1} + \xi_i q_i(\alpha_{i1}^* - \alpha_{i1}) + \xi_i q_i \eta_{i2} \\
&\quad - \lambda_{i1}\xi_i q_i \text{sgn}(\eta_{i1})
\end{aligned} \tag{17}
$$

where $k_{i1}$ and $\lambda_{i1}$ are positive designed parameters. Based on Eqs. (15)-(17), the infinitesimal generator of $V_{i1}$ is given as

$$
\begin{aligned}
\mathcal{L}V_{i1} &= \bar{\zeta}_{i1}^3\Big[\xi_i\Big(q_i(\bar{\zeta}_{i2} + \alpha_{i1} + f_{i1}) - \sum_{m=1}^{N} a_{im}(x_{m2} + f_{m1}) - b_i\dot{y}_r \\
&\quad - \frac{2}{\pi}\arctan(e_{i1}^*)\dot{\sigma}(t)\Big) + (k_{i1}+1)\eta_{i1} + \lambda_{i1}\xi_i q_i \text{sgn}(\eta_{i1})\Big] \\
&\quad + \frac{3}{2}\bar{\zeta}_{i1}^2\xi_i^2\Lambda_i\Lambda_i^T - \frac{1}{\delta_i}\tilde{\Theta}_i\dot{\hat{\Theta}}_i
\end{aligned} \tag{18}
$$

where $\bar{\zeta}_{i2} = \zeta_{i2} - \eta_{i2}$, and $\Lambda_i = q_i g_{i1} - \sum_{m=1}^{N} a_{im}g_{m1}$.

Note that $f_{i1}$ and $f_{m1}$ are unknown functions in Eq. (18). For the sake of subsequent derivation, we sum all terms that contain $f_{i1}$ and $f_{m1}$. The result of the sum is denoted by $\mathscr{F}_{i1}^*$, and the expression for $\mathscr{F}_{i1}^*$ is given as

$$\mathscr{F}_{i1}^* = f_{i1}q_i - \sum_{m=1}^{N} a_{im}(x_{m2} + f_{m1}).$$

To approximate unknown functions $\mathscr{F}_{i1}^*$ and $\Lambda_i\Lambda_i^T$, FLSs $\theta_{i11}^T\Phi_{i11}(X_{i11})$ and $\theta_{i12}^T\Phi_{i12}(X_{i12})$ are utilized, respectively. For any positive constants $\epsilon_{i11}^*$ and $\epsilon_{i12}^*$, one obtains

$$\mathscr{F}_{i1}^* = \theta_{i11}^T\Phi_{i11}(X_{i11}) + \epsilon_{i11}(X_{i11})$$
$$\Lambda_i\Lambda_i^T = \theta_{i12}^T\Phi_{i12}(X_{i12}) + \epsilon_{i12}(X_{i12})$$

where $X_{i11} = [x_{i1}, x_{m1}, x_{m2}]^T$ and $X_{i12} = [x_{i1}, x_{m1}]^T$ ($m \neq i$). $\epsilon_{i11}(X_{i11})$ and $\epsilon_{i12}(X_{i12})$ are approximation errors, and they satisfy $\epsilon_{i11}(X_{i11}) \leq \epsilon_{i11}^*$ and $\epsilon_{i12}(X_{i12}) \leq \epsilon_{i12}^*$, respectively. Based on Lemma 6, it follows that

$$
\begin{aligned}
\bar{\zeta}_{i1}^3\xi_i\mathscr{F}_{i1}^* &= \bar{\zeta}_{i1}^3\xi_i\big[\theta_{i11}^T\Phi_{i11}(X_{i11}) + \epsilon_{i11}(X_{i11})\big] \\
&\leq \frac{3\varepsilon_{i11}^{\frac{4}{3}}}{4}\bar{\zeta}_{i1}^4\xi_i^{\frac{4}{3}}\|\theta_{i11}\|^{\frac{4}{3}}\|\Phi_{i11}\|^{\frac{4}{3}} + \frac{1}{4\varepsilon_{i11}^4} \\
&\quad + \frac{3\varepsilon_{i12}^{\frac{4}{3}}}{4}\bar{\zeta}_{i1}^4\xi_i^{\frac{4}{3}} + \frac{\epsilon_{i11}^{*4}}{4\varepsilon_{i12}^4} \\
&\leq \frac{3\varepsilon_{i11}^{\frac{4}{3}}}{4}\bar{\zeta}_{i1}^4\xi_i^{\frac{4}{3}}\Theta_i\|\Phi_{i11}\|^{\frac{4}{3}} + \frac{1}{4\varepsilon_{i11}^4} \\
&\quad + \frac{3\varepsilon_{i12}^{\frac{4}{3}}}{4}\bar{\zeta}_{i1}^4\xi_i^{\frac{4}{3}} + \frac{\epsilon_{i11}^{*4}}{4\varepsilon_{i12}^4}
\end{aligned} \tag{19}
$$

where $\varepsilon_{i11}$ and $\varepsilon_{i12}$ are positive designed parameters. Similar to inequality (19), the following inequality holds

$$\frac{3}{2}\bar{\zeta}_{i1}^2\xi_i^2\Lambda_i\Lambda_i^T = \frac{3}{2}\bar{\zeta}_{i1}^2\xi_i^2\big[\theta_{i12}^T\Phi_{i12}(X_{i12}) + \epsilon_{i12}(X_{i12})\big]$$

$$
\leq \frac{3\varepsilon_{i13}^2}{4} \bar{\zeta}_{i1}^4 \xi_i^4 \|\theta_{i12}\|^2 \|\Phi_{i12}\|^2 + \frac{3}{4\varepsilon_{i13}^2}
$$

$$
+ \frac{3\varepsilon_{i14}^2}{4} \bar{\zeta}_{i1}^4 \xi_i^4 + \frac{3\epsilon_{i12}^{*2}}{4\varepsilon_{i14}^2}
$$

$$
\leq \frac{3\varepsilon_{i13}^2}{4} \bar{\zeta}_{i1}^4 \xi_i^4 \Theta_i \|\Phi_{i12}\|^2 + \frac{3}{4\varepsilon_{i13}^2}
$$

$$
+ \frac{3\varepsilon_{i14}^2}{4} \bar{\zeta}_{i1}^4 \xi_i^4 + \frac{3\epsilon_{i12}^{*2}}{4\varepsilon_{i14}^2} \tag{20}
$$

where $\varepsilon_{i13}$ and $\varepsilon_{i14}$ are positive designed parameters. By using Lemma 6, it is not difficult to derive

$$
\lambda_{i1} q_i \bar{\zeta}_{i1}^3 \xi_i \mathrm{sgn}(\eta_{i1}) \leq \frac{3}{4} \bar{\zeta}_{i1}^4 \xi_i^{\frac{4}{3}} + \frac{\lambda_{i1}^4 q_i^4}{4}. \tag{21}
$$

By substituting inequalities (19)-(21) to (18), one gets

$$
\mathcal{L}V_{i1} \leq \bar{\zeta}_{i1}^3 \Big[ \xi_i \Big( q_i (\bar{\zeta}_{i2} + \alpha_{i1}) - b_i \dot{y}_r - \frac{2}{\pi} \arctan(e_{i1}^* ) \dot{\sigma}(t)
$$

$$
+ \frac{k_{i1} + 1}{\xi_i} \eta_{i1} \Big) + \frac{3}{4} \bar{\zeta}_{i1} \xi_i^{\frac{1}{3}} + \frac{3\varepsilon_{i11}^{\frac{4}{3}}}{4} \bar{\zeta}_{i1} \xi_i^{\frac{4}{3}} \Theta_i \|\Phi_{i11}\|^{\frac{4}{3}}
$$

$$
+ \frac{3\varepsilon_{i12}^{\frac{4}{3}}}{4} \bar{\zeta}_{i1} \xi_i^{\frac{4}{3}} + \frac{3\varepsilon_{i13}^2}{4} \bar{\zeta}_{i1} \xi_i^4 \Theta_i \|\Phi_{i12}\|^2 + \frac{3\varepsilon_{i14}^2}{4} \bar{\zeta}_{i1} \xi_i^4 \Big]
$$

$$
- \frac{1}{\delta_i} \tilde{\Theta}_i \dot{\hat{\Theta}}_i + \frac{\lambda_{i1}^4 q_i^4}{4} + \frac{1}{4\varepsilon_{i11}^4} + \frac{\epsilon_{i11}^{*4}}{4\varepsilon_{i12}^4} + \frac{3}{4\varepsilon_{i13}^2} + \frac{3\epsilon_{i21}^{*2}}{4\varepsilon_{i14}^2}.
$$

Then, the virtual controller $\alpha_{i1}$ can be designed as

$$
\alpha_{i1} = -\frac{k_{i1} + 1}{\xi_i q_i} \zeta_{i1} + \frac{1}{q_i} \Big( b_i \dot{y}_r + \frac{2}{\pi} \arctan(e_{i1}^*) \dot{\sigma}(t)
$$

$$
- \frac{3\varepsilon_{i11}^{\frac{4}{3}}}{4} \bar{\zeta}_{i1} \xi_i^{\frac{1}{3}} \hat{\Theta}_i \|\Phi_{i11}\|^{\frac{4}{3}} - \frac{3\varepsilon_{i13}^2}{4} \bar{\zeta}_{i1} \xi_i^3 \hat{\Theta}_i \|\Phi_{i12}\|^2
$$

$$
- \frac{3}{4} \bar{\zeta}_{i1} \xi_i^{\frac{1}{3}} - \frac{3\varepsilon_{i12}^{\frac{4}{3}}}{4} \bar{\zeta}_{i1} \xi_i^{\frac{1}{3}} - \frac{3\varepsilon_{i14}^2}{4} \bar{\zeta}_{i1} \xi_i^3 \Big). \tag{22}
$$

By substituting the virtual controller $\alpha_{i1}$ to $\mathcal{L}V_{i1}$, the following inequality holds

$$
\mathcal{L}V_{i1} \leq \bar{\zeta}_{i1}^3 \xi_i q_i \bar{\zeta}_{i2} - (k_{i1} + 1) \bar{\zeta}_{i1}^4 + \frac{1}{\delta_i} \tilde{\Theta}_i (\Delta_{i1} - \dot{\hat{\Theta}}_i) + \Upsilon_{i1}
$$

where $\Delta_{i1} = \frac{3\delta_i \varepsilon_{i11}^{\frac{4}{3}}}{4} \bar{\zeta}_{i1}^4 \xi_i^{\frac{4}{3}} \|\Phi_{i11}\|^{\frac{4}{3}} + \frac{3\delta_i \varepsilon_{i13}^2}{4} \bar{\zeta}_{i1}^4 \xi_i^4 \|\Phi_{i12}\|^2$, and $\Upsilon_{i1} = \frac{\lambda_{i1}^4 q_i^4}{4} + \frac{1}{4\varepsilon_{i11}^4} + \frac{\epsilon_{i11}^{*4}}{4\varepsilon_{i12}^4} + \frac{3}{4\varepsilon_{i13}^2} + \frac{3\epsilon_{i12}^{*2}}{4\varepsilon_{i14}^2}$.

**Step 2**: Construct the following Lyapunov function

$$
V_{i2} = V_{i1} + \frac{1}{4} \bar{\zeta}_{i2}^4
$$

where $\bar{\zeta}_{i2} = \zeta_{i2} - \eta_{i2}$. To eliminate the filtering error $\alpha_{i2}^* - \alpha_{i2}$, the second compensation signal $\eta_{i2}$ is designed as

$$
\dot{\eta}_{i2} = -(k_{i2} + 1)\eta_{i2} + (\alpha_{i2}^* - \alpha_{i2}) - \xi_i q_i \eta_{i1}
$$

$$
+ \eta_{i3} - \lambda_{i2} \mathrm{sgn}(\eta_{i2}) \tag{23}
$$

where $k_{i2}$ and $\lambda_{i2}$ are positive designed parameters.

Similar to Eq. (18), the infinitesimal generator of $V_{i2}$ can be calculated as

$$
\mathcal{L}V_{i2} \leq \bar{\zeta}_{i1}^3 \xi_i q_i \bar{\zeta}_{i2} - (k_{i1} + 1)\bar{\zeta}_{i1}^4 + \frac{1}{\delta_i} \tilde{\Theta}_i (\Delta_{i1} - \dot{\hat{\Theta}}_i) + \Upsilon_{i1}
$$

$$
+ \bar{\zeta}_{i2}^3 \Big( \bar{\zeta}_{i3} + \alpha_{i2} + f_{i2} - \dot{\alpha}_{i1}^* + (k_{i2} + 1)\eta_{i2} + \xi_i q_i \eta_{i1}
$$

$$
+ \lambda_{i2} \mathrm{sgn}(\eta_{i2}) \Big) + \frac{3}{2} \bar{\zeta}_{i2}^2 g_{i2} g_{i2}^T. \tag{24}
$$

Note that unknown nonlinear functions $f_{i2}$ and $g_{i2}g_{i2}^T$ exist in inequality (24). Similar to inequalities (19) and (20), we can obtain

$$
\bar{\zeta}_{i2}^3 f_{i2} \leq \frac{3\varepsilon_{i21}^{\frac{4}{3}}}{4} \bar{\zeta}_{i2}^4 \Theta_i \|\Phi_{i21}\|^{\frac{4}{3}} + \frac{1}{4\varepsilon_{i21}^4} + \frac{3\varepsilon_{i22}^{\frac{4}{3}}}{4} \bar{\zeta}_{i2}^4 + \frac{\epsilon_{i21}^{*4}}{4\varepsilon_{i22}^4}
$$

where $\varepsilon_{i21}$ and $\varepsilon_{i22}$ are positive designed parameters, and $\epsilon_{i21}^*$ is any given positive constant. Moreover, we have

$$
\frac{3}{2} \bar{\zeta}_{i2}^2 g_{i2} g_{i2}^T \leq \frac{3\varepsilon_{i23}^2}{4} \bar{\zeta}_{i2}^4 \Theta_i \|\Phi_{i22}\|^2 + \frac{3}{4\varepsilon_{i23}^2} + \frac{3\varepsilon_{i24}^2}{4} \bar{\zeta}_{i2}^4 + \frac{3\epsilon_{i22}^{*2}}{4\varepsilon_{i24}^2}
$$

where $\varepsilon_{i23}$ and $\varepsilon_{i24}$ are positive designed parameters, and $\epsilon_{i22}^*$ is any given positive constant. By using Lemma 6, it is not difficult to derive

$$
\begin{cases}
\bar{\zeta}_{i2}^3 \xi_i q_i \bar{\zeta}_{i2} \leq \frac{3}{4} \bar{\zeta}_{i1}^4 + \frac{3^3}{4^4} \xi_i^4 q_i^4 \bar{\zeta}_{i2}^4 \\
\bar{\zeta}_{i2}^3 \lambda_{i2} \mathrm{sgn}(\eta_{i2}) \leq \frac{3}{4} \bar{\zeta}_{i2}^4 + \frac{\lambda_{i2}^4}{4}.
\end{cases}
$$

According to the above results, inequality (24) can be further rewritten as

$$
\mathcal{L}V_{i2} \leq \frac{3^3}{4^4} \xi_i^4 q_i^4 \bar{\zeta}_{i2}^4 - k_{i1} \bar{\zeta}_{i1}^4 + \frac{1}{\delta_i} \tilde{\Theta}_i (\Delta_{i1} - \dot{\hat{\Theta}}_i) + \Upsilon_{i1}
$$

$$
+ \bar{\zeta}_{i2}^3 \Big( \bar{\zeta}_{i3} + \alpha_{i2} + \frac{3\varepsilon_{i21}^{\frac{4}{3}}}{4} \bar{\zeta}_{i2} \Theta_i \|\Phi_{i21}\|^{\frac{4}{3}} + \frac{3\varepsilon_{i22}^{\frac{4}{3}}}{4} \bar{\zeta}_{i2}
$$

$$
+ \frac{3\varepsilon_{i23}^2}{4} \bar{\zeta}_{i2} \Theta_i \|\Phi_{i22}\|^2 + \frac{3\varepsilon_{i24}^2}{4} \bar{\zeta}_{i2} - \dot{\alpha}_{i1}^* + \xi_i q_i \eta_{i1}
$$

$$
+ (k_{i2} + 1)\eta_{i2} + \frac{3}{4} \bar{\zeta}_{i2} \Big) + \frac{1}{4\varepsilon_{i21}^4} + \frac{\epsilon_{i21}^{*4}}{4\varepsilon_{i22}^4} + \frac{3}{4\varepsilon_{i23}^2}
$$

$$
+ \frac{3\epsilon_{i22}^{*2}}{4\varepsilon_{i24}^2} + \frac{\lambda_{i2}^4}{4}. \tag{25}
$$

Based on inequality (25), the virtual controller $\alpha_{i2}$ can be constructed as

$$
\alpha_{i2} = -(k_{i2} + 1)\zeta_{i2} + \dot{\alpha}_{i1}^* - \frac{3\varepsilon_{i21}^{\frac{4}{3}}}{4} \bar{\zeta}_{i2} \hat{\Theta}_i \|\Phi_{i21}\|^{\frac{4}{3}}
$$

$$
- \frac{3\varepsilon_{i23}^2}{4} \bar{\zeta}_{i2} \hat{\Theta}_i \|\Phi_{i22}\|^2 - \frac{3\varepsilon_{i22}^{\frac{4}{3}}}{4} \bar{\zeta}_{i2} - \frac{3\varepsilon_{i24}^2}{4} \bar{\zeta}_{i2}
$$

$$
- \frac{3}{4} \bar{\zeta}_{i2} - \xi_i q_i \eta_{i1} - \frac{3^3}{4^4} \xi_i^4 q_i^4 \bar{\zeta}_{i2}. \tag{26}
$$

By substituting Eq. (26) to inequality (25), it obtains

$$
\mathcal{L}V_{i2} \leq \bar{\zeta}_{i2}^3 \bar{\zeta}_{i3} - \sum_{j=1}^{2} k_{ij} \bar{\zeta}_{ij}^4 - \bar{\zeta}_{i2}^4 + \frac{1}{\delta_i} \tilde{\Theta}_i (\Delta_{i2} - \dot{\hat{\Theta}}_i) + \Upsilon_{i2}
$$

where $\Delta_{i2} = \Delta_{i1} + \frac{3\delta_i \varepsilon_{i21}^{\frac{4}{3}}}{4} \bar{\zeta}_{i2}^4 \|\Phi_{i21}\|^{\frac{4}{3}} + \frac{3\delta_i \varepsilon_{i23}^2}{4} \bar{\zeta}_{i2}^4 \|\Phi_{i22}\|^2$, and $\Upsilon_{i2} = \Upsilon_{i1} + \frac{1}{4\varepsilon_{i21}^4} + \frac{\epsilon_{i21}^{*4}}{4\varepsilon_{i22}^4} + \frac{3}{4\varepsilon_{i23}^2} + \frac{3\epsilon_{i22}^{*2}}{4\varepsilon_{i24}^2} + \frac{\lambda_{i2}^4}{4}$.

**Step $o$ $(2 < o < n_i)$**: Choose the Lyapunov function as

$$
V_{io} = V_{i(o-1)} + \frac{1}{4} \bar{\zeta}_{io}^4
$$

where $\bar{\zeta}_{io} = \zeta_{io} - \eta_{io}$. Similar to Eq. (23), to eliminate the filtering error $\alpha_{io}^* - \alpha_{io}$, the $o$-th compensation signal $\eta_{io}$ is designed as

$$\dot{\eta}_{io} = -(k_{io}+1)\,\eta_{io} + (\alpha_{io}^* - \alpha_{io}) - \eta_{i(o-1)}$$
$$+ \eta_{i(o+1)} - \lambda_{io}\mathrm{sgn}(\eta_{io})$$

where $k_{io}$ and $\lambda_{io}$ are positive designed parameters.

The infinitesimal generator of $V_{io}$ satisfies

$$\mathcal{L}V_{io} \leq \bar{\zeta}_{io}^3\Big[\zeta_{i(o+1)} + \alpha_{io} + f_{io} - \dot{\alpha}_{i(o-1)}^* + (k_{io}+1)\,\eta_{io}$$
$$+ \eta_{i(o-1)} + \lambda_{io}\mathrm{sgn}(\eta_{io})\Big] + \frac{3}{2}\bar{\zeta}_{io}^2 g_{io}g_{io}^T + \bar{\zeta}_{i(o-1)}^3\bar{\zeta}_{io}$$
$$- \sum_{j=1}^{o-1} k_{ij}\tilde{\zeta}_{ij}^4 - \bar{\zeta}_{io}^4 + \frac{1}{\delta_i}\tilde{\Theta}_i\Big[\Delta_{i(o-1)} - \dot{\hat{\Theta}}_i\Big] + \Upsilon_{i(o-1)}$$

where $f_{io}$ and $g_{io}g_{io}^T$ are unknown nonlinear functions. Similar to the relationships (19) and (20), we directly present the processing result as follows

$$\bar{\zeta}_{io}^3 f_{io} \leq \frac{3\varepsilon_{io1}^{\frac{4}{3}}}{4}\tilde{\zeta}_{io}^4\Theta_i\|\Phi_{io1}\|^{\frac{4}{3}} + \frac{1}{4\varepsilon_{io1}^4} + \frac{3\varepsilon_{io2}^{\frac{4}{3}}}{4}\tilde{\zeta}_{io}^4 + \frac{\epsilon_{io1}^{*4}}{4\varepsilon_{io2}^4}$$

where $\varepsilon_{io1}$ and $\varepsilon_{io2}$ are positive designed parameters. $\epsilon_{io1}^*$ is any given positive constant. Meanwhile, the following inequality can be obtained

$$\frac{3}{2}\bar{\zeta}_{io}^2 g_{io}g_{io}^T \leq \frac{3\varepsilon_{io3}^2}{4}\tilde{\zeta}_{io}^4\Theta_i\|\Phi_{io2}\|^2 + \frac{3}{4\varepsilon_{io3}^2} + \frac{3\varepsilon_{io4}^2}{4}\tilde{\zeta}_{io}^4 + \frac{3\epsilon_{io2}^{*2}}{4\varepsilon_{io4}^2}$$

where $\varepsilon_{io3}$ and $\varepsilon_{io4}$ are positive designed parameters, and $\epsilon_{io2}^*$ is any given positive constant. By using Lemma 6, we have

$$\begin{cases} \bar{\zeta}_{i(o-1)}^3\bar{\zeta}_{io} \leq \bar{\zeta}_{i(o-1)}^4 + \frac{3^3}{4^4}\tilde{\zeta}_{io}^4 \\ \bar{\zeta}_{io}^3\lambda_{io}\mathrm{sgn}(\eta_{io}) \leq \frac{3}{4}\tilde{\zeta}_{io}^4 + \frac{\lambda_{io}^4}{4}. \end{cases}$$

Based on the above inequalities, $\mathcal{L}V_{io}$ further satisfies the following inequality

$$\mathcal{L}V_{io} \leq \bar{\zeta}_{io}^3\Big(\bar{\zeta}_{i(o+1)} + \alpha_{io} + \frac{3\varepsilon_{io1}^{\frac{4}{3}}}{4}\bar{\zeta}_{io}\Theta_i\|\Phi_{io1}\|^{\frac{4}{3}} + \frac{3\varepsilon_{io2}^{\frac{4}{3}}}{4}\bar{\zeta}_{io}$$
$$+ \frac{3\varepsilon_{io3}^2}{4}\bar{\zeta}_{io}\Theta_i\|\Phi_{io2}\|^2 + \frac{3\varepsilon_{io4}^2}{4}\bar{\zeta}_{io} - \dot{\alpha}_{i(o-1)}^* + \frac{3}{4}\bar{\zeta}_{io}$$
$$+ \eta_{i(o-1)} + (k_{io}+1)\eta_{io} + \frac{3^3}{4^4}\bar{\zeta}_{io}\Big) + \frac{1}{4\varepsilon_{io1}^4}$$
$$+ \frac{\epsilon_{io1}^{*4}}{4\varepsilon_{io2}^4} + \frac{3}{4\varepsilon_{io3}^2} + \frac{3\epsilon_{io2}^{*2}}{4\varepsilon_{io4}^2} + \frac{\lambda_{io}^4}{4} - \sum_{j=1}^{o-1} k_{ij}\tilde{\zeta}_{ij}^4$$
$$+ \frac{1}{\delta_i}\tilde{\Theta}_i\Big[\Delta_{i(o-1)} - \dot{\hat{\Theta}}_i\Big] + \Upsilon_{i(o-1)}. \quad (27)$$

According to inequality (27), the virtual controller $\alpha_{io}$ can be constructed as

$$\alpha_{io} = -(k_{io}+1)\,\bar{\zeta}_{io} + \dot{\alpha}_{i(o-1)}^* - \frac{3\varepsilon_{io1}^{\frac{4}{3}}}{4}\bar{\zeta}_{io}\hat{\Theta}_i\|\Phi_{io1}\|^{\frac{4}{3}}$$
$$- \frac{3\varepsilon_{io3}^2}{4}\bar{\zeta}_{io}\hat{\Theta}_i\|\Phi_{io2}\|^2 - \frac{3\varepsilon_{io2}^{\frac{4}{3}}}{4}\bar{\zeta}_{io} - \frac{3\varepsilon_{io4}^2}{4}\bar{\zeta}_{io}$$
$$- \frac{3}{4}\bar{\zeta}_{io} - \frac{3^3}{4^4}\bar{\zeta}_{io} - \eta_{i(o-1)}. \quad (28)$$

By substituting Eq. (28) to inequality (27), it gets

$$\mathcal{L}V_{io} \leq \bar{\zeta}_{io}^3\bar{\zeta}_{i(o+1)} - \sum_{j=1}^{o} k_{ij}\tilde{\zeta}_{ij}^4 - \bar{\zeta}_{io}^4 + \frac{1}{\delta_i}\tilde{\Theta}_i\big(\Delta_i - \dot{\hat{\Theta}}_i\big) + \Upsilon_{io}$$

where $\Delta_{io} = \Delta_{i(o-1)} + \frac{3\delta_i\varepsilon_{io1}^{\frac{4}{3}}}{4}\tilde{\zeta}_{io}^4\|\Phi_{io1}\|^{\frac{4}{3}} + \frac{3\delta_i\varepsilon_{io3}^2}{4}\tilde{\zeta}_{io}^4\|\Phi_{io2}\|^2$, and $\Upsilon_{io} = \Upsilon_{i(o-1)} + \frac{1}{4\varepsilon_{io1}^4} + \frac{\epsilon_{io1}^{*4}}{4\varepsilon_{io2}^4} + \frac{3}{4\varepsilon_{io3}^2} + \frac{3\epsilon_{io2}^{*2}}{4\varepsilon_{io4}^2} + \frac{\lambda_{io}^4}{4}$.

**Step $n_i$:** In this section, the fault-tolerant controller $u_{ih}$ will be constructed. For subsequent stability analysis of the controller, the compensation signal $\eta_{in_i}$ is defined as

$$\dot{\eta}_{in_i} = -(k_{in_i}+1)\,\eta_{in_i} - \eta_{i(n_i-1)} - \lambda_{in_i}\mathrm{sgn}(\eta_{in_i})$$

where $k_{in_i}$ and $\lambda_{in_i}$ are positive designed parameters.

Based on Assumption 2, it is not difficult to conclude $\inf \sum_{h=1}^{M} |l_{ih}|\rho_{ih_t} \geq \min\{|l_{i1}|_{\rho_{i1_t}}, \ldots, |l_{iM}|_{\rho_{iM_t}}\} > 0$.

To construct subsequent Lyapunov function $V_{in_i}$, unknown constants $\varrho_i$ and $\vartheta_i$ are defined as

$$\varrho_i = \inf_{t\geq 0}\sum_{h=1}^{M}|l_{ih}|\rho_{ih_t}, \ \varphi_i = \frac{1}{\varrho_i}$$

$$\vartheta_i = \sup_{t\geq 0}\sum_{h=1}^{M}|l_{ih}|\nu_{ih_t}.$$

The Lyapunov function is selected as

$$V_{in_i} = V_{i(n_i-1)} + \frac{1}{4}\bar{\zeta}_{in_i}^4 + \frac{\varrho_i}{2\Psi_i}\tilde{\varphi}_i^2 + \frac{1}{2\Gamma_i}\tilde{\vartheta}_i^2$$

where $\Psi_i$ and $\Gamma_i$ are positive design parameters, and $\bar{\zeta}_{in_i} = \zeta_{in_i} - \eta_{in_i}$ is a compensated error.

Similar to Eq. (24), the infinitesimal generator of $V_{in_i}$ satisfies the following relationship

$$\mathcal{L}V_{in_i} \leq \bar{\zeta}_{i(n_i-1)}^3\bar{\zeta}_{in_i} - \sum_{j=1}^{n_i-1} k_{ij}\tilde{\zeta}_{ij}^4 - \bar{\zeta}_{i(n_i-1)}^4 + \Upsilon_{i(n_i-1)}$$
$$+ \frac{1}{\delta_i}\tilde{\Theta}_i(\Delta_{i(n_i-1)} - \dot{\hat{\Theta}}_i) + \frac{3}{2}\bar{\zeta}_{in_i}^2 g_{in_i}g_{in_i}^T$$
$$+ \bar{\zeta}_{in_i}^3\Big(\sum_{h=1}^{M} l_{ih}(\rho_{ih_t}u_{ih} + \nu_{ih_t}) + f_{in_i} - \dot{\alpha}_{i(n_i-1)}^*$$
$$+ (k_{in_i}+1)\,\eta_{in_i} + \eta_{i(n_i-1)} + \lambda_{in_i}\mathrm{sgn}(\eta_{in_i})\Big)$$
$$- \frac{\varrho_i}{\Psi_i}\tilde{\varphi}_i\varepsilon\dot{\hat{\varphi}}_i - \frac{1}{\Gamma_i}\tilde{\vartheta}_i\dot{\hat{\vartheta}}_t \quad (29)$$

where $f_{in_i}$ and $g_{in_i}g_{in_i}^T$ are unknown functions. Similar to the relationships (19) and (20), one has

$$\bar{\zeta}_{in_i}^3 f_{in_i} \leq \frac{3\varepsilon_{in_i1}^{\frac{4}{3}}}{4}\tilde{\zeta}_{in_i}^4\Theta_i\|\Phi_{in_i1}\|^{\frac{4}{3}} + \frac{1}{4\varepsilon_{in_i1}^4} + \frac{3\varepsilon_{in_i2}^{\frac{4}{3}}}{4}\tilde{\zeta}_{in_i}^4 + \frac{\epsilon_{in_i1}^{*4}}{4\varepsilon_{in_i2}^4}$$

where $\varepsilon_{in_i1}$ and $\varepsilon_{in_i2}$ are positive designed parameters. $\epsilon_{in_i1}^*$ is any given positive constant. At the same time, we have

$$\frac{3}{2}\bar{\zeta}_{in_i}^2 g_{in_i}g_{in_i}^T \leq \frac{3\varepsilon_{in_i3}^2}{4}\tilde{\zeta}_{in_i}^4\Theta_i\|\Phi_{in_i1}\|^2 + \frac{3}{4\varepsilon_{in_i3}^2}$$
$$+ \frac{3\varepsilon_{in_i4}^2}{4}\tilde{\zeta}_{in_i}^4 + \frac{3\epsilon_{in_i2}^{*2}}{4\varepsilon_{in_i4}^2}$$

where $\varepsilon_{in_i3}$ and $\varepsilon_{in_i4}$ are positive designed parameters. The scalar $\epsilon_{in_i2}^*$ is a positive constant. By Lemma 6, one gets

$$\begin{cases} \bar{\zeta}_{i(n_i-1)}^3 \bar{\zeta}_{in_i} \leq \bar{\zeta}_{i(n_i-1)}^4 + \dfrac{3^3}{4^4} \bar{\zeta}_{in_i}^4 \\ \bar{\zeta}_{in_i}^3 \lambda_{in_i} \mathrm{sgn}(\eta_{in_i}) \leq \bar{\zeta}_{in_i}^4 + \dfrac{3^3}{4^4} \lambda_{in_i}^4. \end{cases}$$

By substituting the above inequalities to (29), one has

$$\begin{aligned} \mathcal{L}V_{in_i} \leq & -\sum_{j=1}^{n_i-1} k_{ij}\bar{\zeta}_{ij}^4 + \frac{1}{\delta_i}\tilde{\Theta}_i(\Delta_{i(n_i-1)} - \dot{\hat{\Theta}}_i) + \Upsilon_{i(n_i-1)} \\ & + \bar{\zeta}_{in_i}^3 \Big( \sum_{h=1}^{M} l_{ih}\rho_{ih_i}u_{ih} + \sum_{h=1}^{M} l_{ih}\nu_{ih_i} + \bar{u}_i - \bar{u}_i \\ & + \frac{3\varepsilon_{in_i1}^{\frac{4}{3}}}{4}\bar{\zeta}_{in_i}\Theta_i\|\Phi_{in_i1}\|^{\frac{4}{3}} + \frac{3\varepsilon_{in_i3}^2}{4}\bar{\zeta}_{in_i}\Theta_i\|\Phi_{in_i2}\|^2 \\ & + \frac{3\varepsilon_{in_i2}^{\frac{4}{3}}}{4}\bar{\zeta}_{in_i} + \frac{3\varepsilon_{in_i4}^2}{4}\bar{\zeta}_{in_i} + \frac{3^3}{4^4}\bar{\zeta}_{in_i} + \bar{\zeta}_{in_i} - \dot{\alpha}_{i(n_i-1)}^* \\ & + (k_{in_i}+1)\eta_{in_i} + \eta_{i(n_i-1)} \Big) - \frac{\varrho_i}{\Psi_i}\bar{\varphi}_i\dot{\hat{\varphi}}_i - \frac{1}{\Gamma_i}\tilde{\vartheta}_i\dot{\hat{\vartheta}}_i \\ & + \frac{3^3}{4^4}\lambda_{in_i}^4 + \frac{1}{4\varepsilon_{in_i1}^4} + \frac{\epsilon_{in_i1}^{*4}}{4\varepsilon_{in_i2}^4} + \frac{3}{4\varepsilon_{in_i3}^2} + \frac{3\epsilon_{in_i2}^{*2}}{4\varepsilon_{in_i4}^2} \quad (30) \end{aligned}$$

where $\bar{u}_i$ is an intermediate control variable described by

$$\begin{aligned} \bar{u}_i = & (k_{in_i}+1)\zeta_{in_i} + \frac{3\varepsilon_{in_i1}^{\frac{4}{3}}}{4}\bar{\zeta}_{in_i}\hat{\Theta}_i\|\Phi_{in_i1}\|^{\frac{4}{3}} + \frac{3\varepsilon_{in_i2}^{\frac{4}{3}}}{4}\bar{\zeta}_{in_i} \\ & + \frac{3\varepsilon_{in_i3}^2}{4}\bar{\zeta}_{in_i}\hat{\Theta}_i\|\Phi_{in_i2}\|^2 + \frac{3\varepsilon_{in_i4}^2}{4}\bar{\zeta}_{in_i} + \frac{3^3}{4^4}\bar{\zeta}_{in_i} \\ & - \dot{\alpha}_{i(n_i-1)}^* + \eta_{i(n_i-1)} + \hat{\vartheta}_i \tanh\left(\frac{\bar{\zeta}_{in_i}^3}{\epsilon_i}\right) \quad (31) \end{aligned}$$

and $\epsilon_i$ is a positive designed parameter. By substituting Eq. (31) to inequality (30), one gets

$$\begin{aligned} \mathcal{L}V_{in_i} \leq & -\sum_{j=1}^{n_i} k_{ij}\bar{\zeta}_{ij}^4 + \frac{1}{\delta_i}\tilde{\Theta}_i(\Delta_{in_i} - \dot{\hat{\Theta}}_i) + \Upsilon_{i(n_i-1)} \\ & + \bar{\zeta}_{in_i}^3 \sum_{h=1}^{M} l_{ih}\rho_{ih_i}u_{ih} + \bar{\zeta}_{in_i}^3\bar{u}_i + \frac{3^3}{4^4}\lambda_{in_i}^4 \\ & - \frac{\varrho_i}{\Psi_i}\bar{\varphi}_i\dot{\hat{\varphi}}_i - \frac{1}{\Gamma_i}\tilde{\vartheta}_i\Big(\dot{\hat{\vartheta}}_i - \Gamma_i\bar{\zeta}_{in_i}^3 \tanh\Big(\frac{\bar{\zeta}_{in_i}^3}{\epsilon_i}\Big)\Big) \\ & + \vartheta_i\Big(|\bar{\zeta}_{in_i}^3| - \bar{\zeta}_{in_i}^3 \tanh\Big(\frac{\bar{\zeta}_{in_i}^3}{\epsilon_i}\Big)\Big) + \frac{1}{4\varepsilon_{in_i1}^4} \\ & + \frac{\epsilon_{in_i1}^{*4}}{4\varepsilon_{in_i2}^4} + \frac{3}{4\varepsilon_{in_i3}^2} + \frac{3\epsilon_{in_i2}^{*2}}{4\varepsilon_{in_i4}^2}. \quad (32) \end{aligned}$$

The variable $\Delta_{in_i}$ satisfies

$$\Delta_{in_i} = \Delta_{i(n_i-1)} + \frac{3\delta_i\varepsilon_{in_i1}^{\frac{4}{3}}}{4}\bar{\zeta}_{in_i}^4\|\Phi_{in_i1}\|^{\frac{4}{3}} + \frac{3\delta_i\varepsilon_{in_i3}^2}{4}\bar{\zeta}_{in_i}^4\|\Phi_{in_i2}\|^2.$$

The fault-tolerant controller $u_{ih}$ is designed as

$$u_{ih} = \mathrm{sgn}(l_{ih})\bar{\alpha}_{in_i}, \quad \bar{\alpha}_{in_i} = -\frac{\bar{\zeta}_{in_i}^3\hat{\varphi}_i^2\bar{u}_i^2}{\sqrt{\bar{\zeta}_{in_i}^6\hat{\varphi}_i^2\bar{u}_i^2 + \varepsilon_{in_i5}^2}} \quad (33)$$

where $\varepsilon_{in_i5}$ is a positive designed parameter. The adaptive laws are constructed as

$$\begin{cases} \dot{\hat{\Theta}}_i = \Delta_{in_i} - \bar{\delta}_i\hat{\Theta}_i \\ \dot{\hat{\vartheta}}_i = \Gamma_i\bar{\zeta}_{in_i}^3 \tanh\left(\dfrac{\bar{\zeta}_{in_i}^3}{\epsilon_i}\right) - \bar{\Gamma}_i\hat{\vartheta}_i \\ \dot{\hat{\varphi}}_i = \Psi_i\bar{\zeta}_{in_i}^3\bar{u}_i - \bar{\Psi}_i\hat{\varphi}_i \end{cases} \quad (34)$$

where $\bar{\delta}_i$, $\bar{\Gamma}_i$ and $\bar{\Psi}_i$ are positive designed parameters. By Lemma 7, it is not hard to derive

$$\begin{aligned} \bar{\zeta}_{in_i}^3 \sum_{h=1}^{M} l_{ih}\rho_{ih_i}u_{ih} &= -\sum_{h=1}^{M} |l_{ih}|\rho_{ih_i}\frac{\bar{\zeta}_{in_i}^6\hat{\varphi}_i^2\bar{u}_i^2}{\sqrt{\bar{\zeta}_{in_i}^6\hat{\varphi}_i^2\bar{u}_i^2 + \varepsilon_{in_i5}^2}} \\ &\leq -\frac{\varrho_i\bar{\zeta}_{in_i}^6\hat{\varphi}_i^2\bar{u}_i^2}{\sqrt{\bar{\zeta}_{in_i}^6\hat{\varphi}_i^2\bar{u}_i^2 + \varepsilon_{in_i5}^2}} \\ &\leq \varrho_i\varepsilon_{in_i5} - \varrho_i\bar{\zeta}_{in_i}^3\hat{\varphi}_i\bar{u}_i. \quad (35) \end{aligned}$$

Based on the fact that $0 \leq |\aleph| - \aleph\tanh(\frac{\aleph}{\bar{\wp}}) \leq 0.2785\bar{\wp}$ and relationships (33)-(35), inequality (32) can be rewritten as

$$\begin{aligned} \mathcal{L}V_{in_i} \leq & -\sum_{j=1}^{n_i} k_{ij}\bar{\zeta}_{ij}^4 + \frac{\bar{\delta}_i}{\delta_i}\tilde{\Theta}_i\hat{\Theta}_i + \frac{\varrho_i\bar{\Psi}_i}{\Psi_i}\bar{\varphi}_i\hat{\varphi}_i + \frac{\bar{\Gamma}_i}{\Gamma_i}\tilde{\vartheta}_i\hat{\vartheta}_i \\ & + \Upsilon_{i(n_i-1)} + \frac{3^3}{4^4}\lambda_{in_i}^4 + \frac{1}{4\varepsilon_{in_i1}^4} + \frac{\epsilon_{in_i1}^{*4}}{4\varepsilon_{in_i2}^4} + \frac{3}{4\varepsilon_{in_i3}^2} \\ & + \frac{3\epsilon_{in_i2}^{*2}}{4\varepsilon_{in_i4}^2} + 0.2785\vartheta_i\bar{\wp}_i + \varrho_i\varepsilon_{in_i5} \quad (36) \end{aligned}$$

where $\bar{\wp}$ is a positive constant, and the variable $\aleph \in R$. By Lemma 6, it is not difficult to get

$$\begin{cases} \dfrac{\bar{\Gamma}_i}{\Gamma_i}\tilde{\vartheta}_i\hat{\vartheta}_i \leq -\dfrac{\bar{\Gamma}_i}{2\Gamma_i}\tilde{\vartheta}_i^2 + \dfrac{\bar{\Gamma}_i}{2\Gamma_i}\vartheta_i^2 \\ \dfrac{\bar{\delta}_i}{\delta_i}\tilde{\Theta}_i\hat{\Theta}_i \leq -\dfrac{\bar{\delta}_i}{2\delta_i}\tilde{\Theta}_i^2 + \dfrac{\bar{\delta}_i}{2\delta_i}\Theta_i^2 \\ \dfrac{\bar{\Psi}_i}{\varrho_i\Psi_i}\bar{\varphi}_i\hat{\varphi}_i \leq -\dfrac{\bar{\Psi}_i}{2\Psi_i}\bar{\varphi}_i^2 + \dfrac{\varrho_i\bar{\Psi}_i}{2\Psi_i}\varphi_i^2. \end{cases}$$

Then inequality (36) can be rewritten as

$$\mathcal{L}V_{in_i} \leq -\sum_{j=1}^{n_i} k_{ij}\bar{\zeta}_{ij}^4 - \frac{\bar{\delta}_i}{2\delta_i}\tilde{\Theta}_i^2 - \frac{\varrho_i\bar{\Psi}_i}{2\Psi_i}\bar{\varphi}_i^2 - \frac{\bar{\Gamma}_i}{2\Gamma_i}\tilde{\vartheta}_i^2 + \Upsilon_{in_i}$$

where $\Upsilon_{in_i} = \Upsilon_{i(n_i-1)} + \frac{3^3}{4^4}\lambda_{in_i}^4 + \frac{1}{4\varepsilon_{in_i1}^4} + \frac{\epsilon_{in_i1}^{*4}}{4\varepsilon_{in_i2}^4} + \frac{3}{4\varepsilon_{in_i3}^2} + \frac{3\epsilon_{in_i2}^{*2}}{4\varepsilon_{in_i4}^2} + 0.2785\vartheta_i\bar{\wp}_i + \varrho_i\varepsilon_{in_i5} + \frac{\bar{\Gamma}_i}{2\Gamma_i}\vartheta_i^2 + \frac{\bar{\delta}_i}{2\delta_i}\Theta_i^2 + \frac{\varrho_i\bar{\Psi}_i}{2\Psi_i}\varphi_i^2$.

*Remark 3:* In the design of the fault-tolerant controller, we adopt two FLSs to deal with unknown nonlinear dynamics in drift and diffusion terms, respectively. Generally speaking, the number of adaptive laws increases as the order of the agent increases, which will lead to the over-parameterization problem. Inspired by [42], a less parameter estimation approach is adopted, which makes the number of adaptive laws independent of the order of the system. Therefore, this approach avoids the problem of over-parameterization and reduces the computational burden in results [43] and [44].

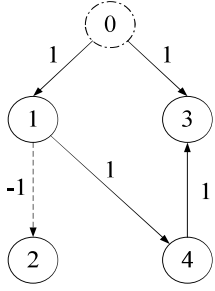

Fig. 1. The communication topology in *Numerical Example*.

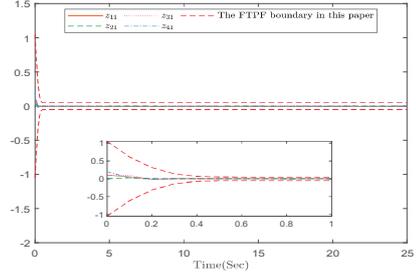

Fig. 3. Bipartite consensus errors $z_{i1}$ under the constraint of the FTPF in this paper.

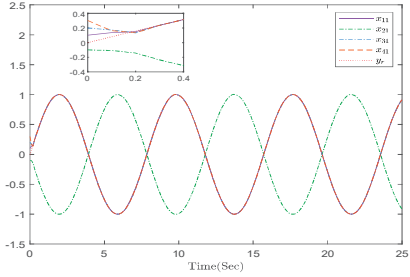

Fig. 2. Output trajectories of agents in *Numerical Example*.

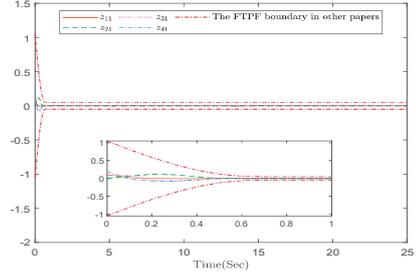

Fig. 4. Bipartite consensus errors $z_{i1}$ under the constraint of the FTPF in [25]–[28].

## V. STABILITY ANALYSIS

*Theorem 1:* For stochastic nonlinear MASs (9) with virtual controllers (22) and (28), the fault-tolerant controller (33) and adaptive laws (34), if *Assumptions* 1-3 and the condition $|z_{i1}(0)| \in (0, \sigma(0))$ are satisfied, then *control objectives* (a) and (b) can be achieved.

*Proof:* Firstly, to prove that the compensation signal $\eta_{ij}$ is bounded, the following Lyapunov function is considered

$$V_{i\eta} = \sum_{j=1}^{n_i} \frac{\eta_{ij}^2}{2}.$$

Then, it obtains

$$
\begin{aligned}
\dot{V}_{i\eta} = & -(k_{i1}+1)\eta_{i1}^2 + \xi_i q_i (\alpha_{i1}^* - \alpha_{i1})\eta_{i1} \\
& + \xi_i q_i \eta_{i1} \eta_{i2} - \lambda_{i1} \xi_i q_i |\eta_{i1}| \\
& - (k_{i2}+1)\eta_{i2}^2 + (\alpha_{i2}^* - \alpha_{i2})\eta_{i2} - \xi_i q_i \eta_{i1} \eta_{i2} \\
& + \eta_{i2}\eta_{i3} - \lambda_{i2}|\eta_{i2}| \\
& \cdots \\
& - (k_{in_i}+1)\eta_{in_i}^2 - \eta_{i(n_i-1)}\eta_{in_i} - \lambda_{in_i}|\eta_{in_i}|.
\end{aligned}
\tag{37}
$$

According to the results in [45], the filter error satisfies the condition that $\|\alpha_{ij}^* - \alpha_{ij}\| \le \sigma_{ij}$ in a finite time with a known constant $\sigma_{ij}$. Then, Eq. (37) can be rewritten

$$\dot{V}_{i\eta} \le -\sum_{j=1}^{n_i}(k_{ij}+1)\eta_{ij}^2 + \sum_{j=1}^{n_i}\bar{l}_j \sigma_{ij}|\eta_{ij}| - \sum_{j=1}^{n_i}\bar{l}_j \lambda_{ij}|\eta_{ij}|$$

where $\bar{l}_1 = \xi_i q_i$, and $\bar{l}_j = 1$ with $j = 2, 3, \ldots, n_i$. By selecting appropriate parameters so that $\sigma_{ij}$ and $\lambda_{ij}$ satisfy the condition that $\sigma_{ij} < \lambda_{ij}$, it gets

$$\dot{V}_{i\eta} \le -\sum_{j=1}^{n_i}(k_{ij}+1)\eta_{ij}^2$$

which indicates that the compensation signal $\eta_{ij}$ can exponentially converge to 0.

Construct the following Lyapunov function

$$V = \sum_{i=1}^{N} V_{in_i}.$$

Then, one obtains

$$
\begin{aligned}
\mathcal{L}V \le & \sum_{i=1}^{N}\Big( -\sum_{j=1}^{n_i} k_{ij}\bar{\zeta}_{ij}^4 - \frac{\bar{\delta}_i}{2\bar{\delta}_i}\tilde{\Theta}_i^2 - \frac{\varrho_i \bar{\Psi}_i}{2\Psi_i}\tilde{\varphi}_i^2 - \frac{\bar{\Gamma}_i}{2\Gamma_i}\tilde{\vartheta}_i^2 + \Upsilon_{in_i} \Big) \\
\le & -cV + b
\end{aligned}
\tag{38}
$$

where $c = \min\{4k_{ij}, \bar{\delta}_i, \bar{\Psi}_i, \bar{\Gamma}_i\}$, $b = \Sigma_{i=1}^{N}\Upsilon_{in_i}$. From Eq. (38) and Lemma 3, it gets

$$E(V) \le V(0)e^{-ct} + \frac{b}{c}.
\tag{39}$$

According to the definition of Lyapunov function $V$, inequality (39) indicates that signals $\bar{\zeta}_{ij}$, $\tilde{\Theta}_i$, $\tilde{\varphi}_i$ and $\tilde{\vartheta}_i$ are all SGUUB in probability. Based on the convergence of compensation signal $\eta_{ij}$ and the relationship $\zeta_{ij} = \bar{\zeta}_{ij} + \eta_{ij}$, it can be obtained that the signal $\zeta_{ij}$ is also SGUUB in probability. Furthermore, by considering the error transformation mechanism (12), the coordinate transformations (16), virtual control laws (22) and (28), and the fault-tolerant controller (33), we can get that all signals of agents are SGUUB in probability. The *control objective* (a) is achieved.

In addition, based on Eq. (12) and the fourth property of FTPF, it follows that

$$z_{i1} = \sigma_\infty \mu(e_{i1}^*), \quad \forall t \ge T_s$$

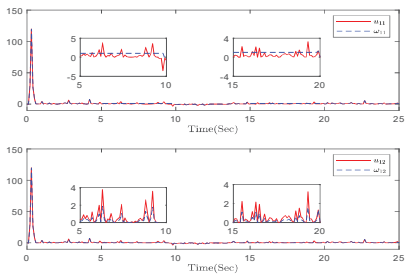

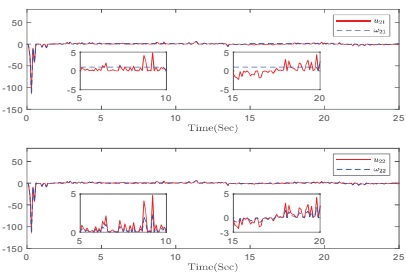

Fig. 5. The curves of actuator inputs and outputs of the first follower.

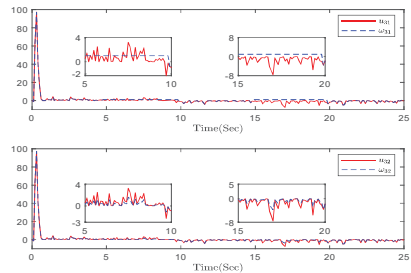

Fig. 7. The curves of actuator inputs and outputs of the third follower.

Fig. 6. The curves of actuator inputs and outputs of the second follower.

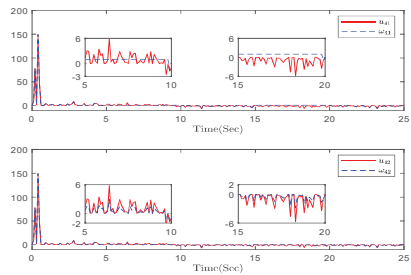

Fig. 8. The curves of actuator inputs and outputs of the fourth follower.

which indicates that the relationship $|z_{i1}| < \sigma_\infty$ holds. Then, for $\forall t \geq T_s$, we can further yield $E(\|z_{i1}\|) < \sigma_\infty$, which implies the *control objective* (b) is satisfied.

Moreover, define bipartite tracking errors $\breve{e}_i = y_i - \Im_i y_r$ with $i = 1, 2, \ldots, N$, and vectors $\breve{e} = [\breve{e}_1, \ldots, \breve{e}_N]^T$ and $z = [z_{11}, \ldots, z_{N1}]^T$. Based on Lemma 1 and the fact that $z = (L_{\mathscr{G}} + B_{\mathscr{G}})\breve{e}$, we have

$$\|\breve{e}\| \leq \frac{\|z\|}{\hbar(L_{\mathscr{G}} + B_{\mathscr{G}})} \tag{40}$$

where $\hbar(L_{\mathscr{G}} + B_{\mathscr{G}})$ represents the minimum singular value of the matrix $(L_{\mathscr{G}} + B_{\mathscr{G}})$. Based on inequality (40) and the fact $|z_{i1}| < \sigma_\infty$, the following relationship holds

$$E(\|\breve{e}\|) \leq \frac{E(\|z\|)}{\hbar(L_{\mathscr{G}} + B_{\mathscr{G}})} < \frac{\sigma_\infty \sqrt{N}}{\hbar(L_{\mathscr{G}} + B_{\mathscr{G}})}, \quad \forall t \geq T_s. \tag{41}$$

Due to the fact that $\sigma_\infty$ can be arbitrarily small, inequality (41) indicates that the tracking error can also satisfy the arbitrary tracking precision with probability in the predefined settling time. The proof is completed. ∎

## VI. Simulation Results

In this section, to verify its effectiveness, the proposed control scheme is applied to BCT tasks of second-order stochastic MASs and a group of vehicles, respectively.

*Numerical Example:* Consider MASs consisting of four followers (agents 1-4) and a leader (agent 0). The communication topology of MASs is denotes as Fig. 1. Based on Fig. 1, we can obtain that $\hbar(L_{\mathscr{G}} + B_{\mathscr{G}}) = 1$. Let $y_r = \frac{30}{9}\sin(\frac{4t}{5})$ denote the leader signal. The $i$-th follower is modeled as

$$\begin{cases} dx_{i1} = (x_{i2} + 0.2x_{i1})dt + 0.2\sin(6x_{i1})d\varpi \\ dx_{i2} = \left(\sum_{h=1}^{2} l_{ih}\omega_{ih} + 0.2x_{i1}x_{i2}\right)dt + 0.2\sin(6x_{i1}x_{i2})d\varpi \\ y_i = x_{i1}, \quad i = 1, 2, 3, 4. \end{cases}$$

The $i$-th follower suffers from the following actuator faults

$$\begin{cases} \omega_{i1}(t) = \nu_{i1\iota}, \quad t \in [5\iota, 5(\iota+1)) \\ \omega_{i2}(t) = \rho_{i2\iota}u_{i2}(t), \quad \iota = 1, 3, \ldots \end{cases} \tag{42}$$

where $\nu_{i1\iota} = 1$ and $\rho_{i2\iota} = 0.5$. Eq. (42) indicates that the first actuator operates in TLOE every 5 seconds, and the second actuator operates in PLOE every 5 seconds.

The designed parameters of $\sigma(t)$ are selected as $\sigma_0 = 1.05$, $\sigma_\infty = 0.05$, $\varsigma = 4$ and $T_s = 0.8$. The first fuzzy membership functions are chosen as $\phi_{g_i}^{\iota}(x_i) = e^{-0.5(x_i - \bar{\chi}_\iota)^2}$ with $\iota = 1, 2, \ldots, 7$, where $\bar{\chi}_1 = -1.5$, $\bar{\chi}_2 = -1$, $\bar{\chi}_3 = -0.5$, $\bar{\chi}_4 = 0$, $\bar{\chi}_5 = 0.5$, $\bar{\chi}_6 = 1$, $\bar{\chi}_7 = 1.5$. The second fuzzy membership functions are chosen as $\phi_{g_i}^{\iota}(x_i) = e^{-0.5(x_i - \bar{\chi}_\iota)^2}$ with $\iota = 1, 2, \ldots, 7$, where $\bar{\chi}_1 = -2$, $\bar{\chi}_2 = -1.5$, $\bar{\chi}_3 = -0.5$, $\bar{\chi}_4 = 0$, $\bar{\chi}_5 = 0.5$, $\bar{\chi}_6 = 1.5$, $\bar{\chi}_7 = 2$. The initial states are set as $x_{11}(0) = 0.1$, $x_{21}(0) = -0.1$, $x_{31}(0) = 0.2$, $x_{41}(0) = 0.3$ and $x_{i2}(0) = 0$, $\hat{\Theta}_i(0) = 1$, $\hat{\vartheta}_i(0) = 1$ and $\hat{\varphi}_i(0) = 1$. The designed parameters of the controller are set as $k_{11} = k_{21} = k_{12} = k_{22} = 4$, $k_{31} = k_{41} = k_{32} = k_{42} = 12$, $l_{i1} = 1$, $l_{i2} = 2$, $\varepsilon_{i11} = \varepsilon_{i21} = 0.5$, $\varepsilon_{i12} = \varepsilon_{i22} = 0.1$, $\varepsilon_{i13} = \varepsilon_{i23} = 0.5$, $\varepsilon_{i14} = \varepsilon_{i24} = 0.1$, $\tau_i = 0.0125$, $\lambda_{i1} = \lambda_{i2} = 0.01$, $\epsilon_i = 0.1$, $\varepsilon_{i25} = 0.1$, $\delta_i = 4$, $\bar{\delta}_i = 0.5$, $\Gamma_i = 4$, $\bar{\Gamma}_i = 0.4$, $\Psi_i = 5$, $\bar{\Psi}_i = 0.2$.

The simulation results of the numerical example are presented in Figs. 2-8. Based on Lemma 2 and Fig. 1, we can obtain the bipartition of nodes $\mathbb{V}_1 = \{1, 3, 4\}$ and $\mathbb{V}_2 = \{2\}$, which means that agents 1, 3, 4 can track the leader signal and agent 2 can track the leader signal in an antisymmetric direction. The results in Fig. 2 verify this inference. Subsequently, in Fig. 3, we give curves for the bipartite consensus errors $z_{i1}$ under the constraint of the FTPF in this paper. From Fig. 3, it can be concluded that $E\left(\|z_{i1}\|\right) < 0.05$ ($\forall t \geq 0.8$). Based on inequality (41), we can further obtain that $E\left(\|\tilde{e}\|\right) < 0.1$ ($\forall t \geq 0.8$). Fig. 4 shows curves for the bipartite consensus errors $z_{i1}$ under the constraint of the FTPF in [25]–[28]. By comparing Fig. 3 and Fig. 4, we can see that the FTPF in this paper improves the transient performance of MASs under the same conditions. Finally, Figs. 5-8 show the curves for the actuator inputs and outputs, and it can be seen that the first and second actuators suffer TLOE and PLOE every 5 seconds, respectively. From the above simulation results, it can be concluded that agents suffering from stochastic disturbances and actuator faults can still complete robust precision BCT tasks in a predefined settling time.

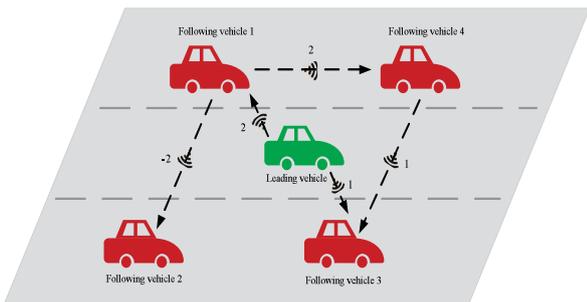

Fig. 9. Communication relationships among vehicles.

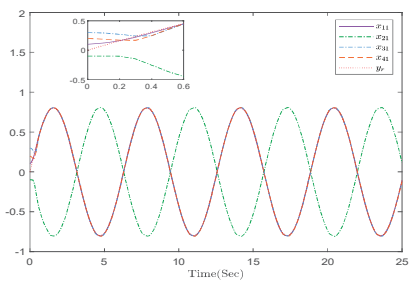

Fig. 10. Output trajectories of agents in *Practical Example*.

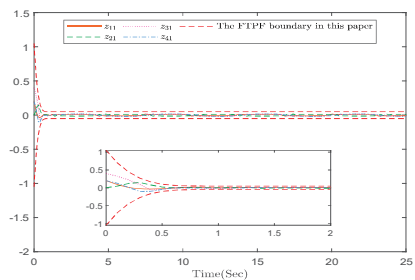

Fig. 11. Bipartite consensus errors $z_{i1}$ under the constraint of the FTPF in this paper.

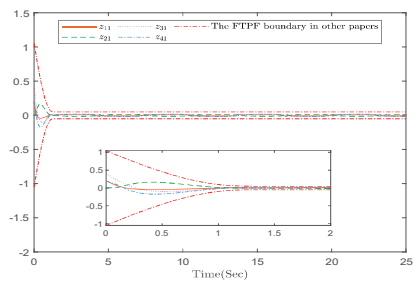

Fig. 12. Bipartite consensus errors $z_{i1}$ under the constraint of the FTPF in [25]–[28].

*Practical Example:* In this example, a group of vehicles [46] travelling on a firm horizontal surface are studied to further illustrate the effectiveness of the proposed control scheme. The communication relationships among vehicles are described in Fig. 9. Based on Fig. 9, we can calculate that $\hbar(L_{\mathscr{A}} + B_{\mathscr{A}}) = 2$. Suppose that the FTPF in this paper the trajectory of the leading vehicle is $y_r = 0.8\sin(t)$. The $i$-th following vehicle is modeled as

$$m_i \ddot{q}_i + \kappa_i m_i g + \varpi_i \dot{q}_i = \sum_{h=1}^{2} \omega_{ih}, \quad i = 1, 2, 3, 4, \qquad (43)$$

where $\dot{q}_i$ and $\ddot{q}_i$ represent the velocity and acceleration of the following vehicle, respectively. $m_i$ denotes the mass of the vehicle. $\kappa_i$ is the kinetic friction coefficient. $g$ is the acceleration of gravity. $\varpi_i \dot{q}_i$ is a viscous friction with an unknown constant $\varpi$. Assume that $\varpi_i = \hat{\varpi}_i + \mathcal{S}_i W_t$, where $\hat{\varpi}_i$ and $\mathcal{S}_i$ are constants, and $W_t$ is a standard white noise process. Define state variables $x_{i1} = q_i$, $x_{i2} = \dot{q}$, Eq. (43) can be rewritten as

$$
\begin{cases}
dx_{i1} = x_{i2}dt \\
dx_{i2} = \left(\sum_{h=1}^{2} \dfrac{1}{m_i}\omega_{ih} - \dfrac{\hat{\varpi}_i}{m_i}x_{i2} - \kappa_i g\right) dt + \dfrac{\mathcal{S}_i}{m_i}d\varpi \\
y_i = x_{i1}
\end{cases}
$$

where $m_i = 0.5$ kg, $g = 10$ m/s$^2$, $\kappa_i = 0.02$, $\hat{\varpi}_i = 0.5$, $\mathcal{S}_i = 0.1$, and $\omega_{ih}$ ($h = 1, 2$) are the same as Eq. (42).

The designed parameters of $\sigma(t)$ are selected as $\sigma_0 = 1.05$, $\sigma_\infty = 0.05$, $\varsigma = 6$ and $T_s = 1.5$. Based on inequality (41), with the help of the FTPF, the inequality $E\left(\|\tilde{e}\|\right) < 0.05$ ($\forall t \geq 1.5$) holds, which indicates that the tracking precision of vehicles is set as $\pm 0.05$ meter.

The fuzzy membership functions are the same as the numerical example. The initial states of vehicles are set as $x_{11}(0) = 0.1$, $x_{21}(0) = -0.1$, $x_{31}(0) = 0.3$, $x_{41}(0) = 0.2$ and $x_{i2}(0) = 0$. The initial values of adaptive laws are chosen as $\hat{\Theta}_i(0) = 1$, $\hat{\vartheta}_i(0) = 1$ and $\hat{\varphi}_i(0) = 1$. The designed parameters are set as $k_{i1} = k_{i2} = 5$, $\varepsilon_{i11} = \varepsilon_{i21} = 0.5$, $\varepsilon_{i12} = \varepsilon_{i22} = 0.1$, $\varepsilon_{i13} = \varepsilon_{i23} = 0.5$, $\varepsilon_{i14} = \varepsilon_{i24} = 0.1$, $\tau_i = 0.014$, $\lambda_{i1} = \lambda_{i2} = 0.01$, $\epsilon_i = 0.2$, $\varepsilon_{i25} = 0.2$, $\delta_i = 1$, $\bar{\delta}_i = 0.4$, $\Gamma_i = 1$, $\bar{\Gamma}_i = 0.4$, $\Psi_i = 1$, $\bar{\Psi}_i = 0.8$.

The simulation results of the practical example are presented in Figs. 10-16. The bipartite tracking trajectories of vehicles are shown in Fig. 10. Under the constraint of the FTPF in this paper, bipartite consensus errors $z_{i1}$ are presented in Fig. 11. Fig. 12 shows curves for bipartite consensus errors $z_{i1}$ under the constraint of the FTPF in [25]–[28]. From Fig. 11 and Fig. 12, it can be concluded that $E\left(\|z_{i1}\|\right) < 0.05$ ($\forall t \geq 1.5$). Based on inequality (41), we can further obtain that $E\left(\|\check{e}\|\right) < 0.05$ ($\forall t \geq 1.5$). By comparing Fig. 11 and Fig. 12, it is not hard to conclude that the FTPF in this paper improves the transient performance of vehicles systems under identical conditions. Finally, Figs. 13-16 show the curves of the actuator inputs and outputs, and it can be seen that the first and second actuators suffer TLOE and PLOE every 5 seconds, respectively. Based on Figs. 10-16, we can conclude that all vehicles subject to stochastic disturbances and actuator faults achieve finite-time robust precision BCT tasks.

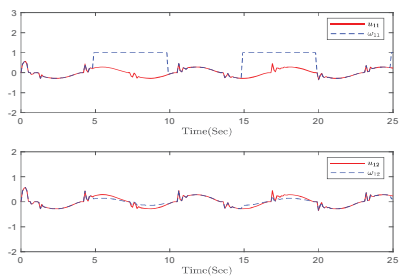

Fig. 13. The curves of actuator inputs and outputs of the first follower.

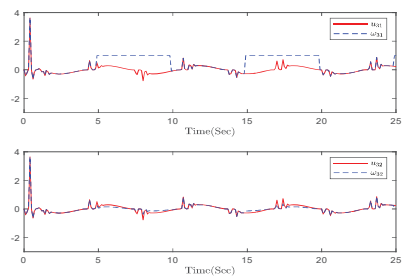

Fig. 15. The curves of actuator inputs and outputs of the third follower.

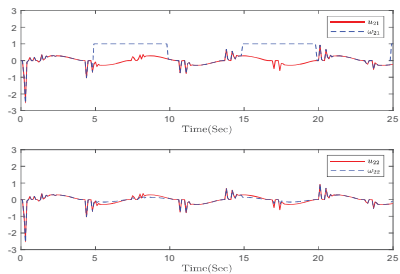

Fig. 14. The curves of actuator inputs and outputs of the second follower.

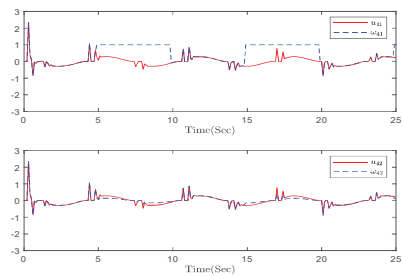

Fig. 16. The curves of actuator inputs and outputs of the fourth follower.

## VII. Conclusions

The presented paper has investigated finite-time robust precision BCT tasks for stochastic nonlinear MASs with actuator faults. An optimization-based FTPF has been proposed to improve the transient performance of MASs. Based on the FTPF, a fuzzy fault-tolerant distributed cooperative control scheme has been developed to ensure that all agents with various uncertainties complete robust precision BCT tasks in the predefined settling time. In addition, the less parameter estimation approach has been employed to avoid the over-parameterization problem in the controller design. Finally, the effectiveness of the proposed scheme has been verified in simulations, and the anticipated performance has been achieved in BCT tasks for a group of vehicles suffering from various uncertainties. For further extensions, we will continue to investigate how to achieve robust precision BCT tasks for MASs with unmeasurable states.